# Electric-field control of the nucleation and motion of isolated three-fold polar vertices

Mingqiang Li[1,2], Tiannan Yang[3], Pan Chen[4], Yongjun Wang[5], Ruixue Zhu[1], Xiaomei Li[1], Ruochen Shi[1], Heng-Jui Liu[6], Yen-Lin Huang[5], Xiumei Ma[1], Jingmin Zhang[1], Xuedong Bai[4,7,8], Long-Qing Chen[3], Ying-Hao Chu[5,9] & Peng Gao[1,10,11,12] ✉

Recently various topological polar structures have been discovered in oxide thin films. Despite the increasing evidence of their switchability under electrical and/or mechanical fields, the dynamic property of isolated ones, which is usually required for applications such as data storage, is still absent. Here, we show the controlled nucleation and motion of isolated three-fold vertices under an applied electric field. At the $PbTiO_3$/$SrRuO_3$ interface, a two-unit-cell thick $SrTiO_3$ layer provides electrical boundary conditions for the formation of three-fold vertices. Utilizing the $SrTiO_3$ layer and in situ electrical testing system, we find that isolated three-fold vertices can move in a controllable and reversible manner with a velocity up to ~629 nm s$^{-1}$. Microstructural evolution of the nucleation and propagation of isolated three-fold vertices is further revealed by phase-field simulations. This work demonstrates the ability to electrically manipulate isolated three-fold vertices, shedding light on the dynamic property of isolated topological polar structures.

A variety of topological polar configurations have been created in complex oxides by precisely mediating the electrical and mechanical boundary conditions[1–3]. For example, vertices (meeting points of two or more domain walls)[4–6], vortices (require a non-zero polarization curl)[7], polar skyrmions[8], and polar merons[9] have been synthesized in $PbTiO_3$ films, $(PbTiO_3)_n$/$(SrTiO_3)_n$ superlattices or directly written in ferroelectrics by scanning probe techniques. These topological structures host unique properties that allow the development of novel electronics, including negative capacitance field-effect transistors[10,11] and high-density nonvolatile memories[12].

Practical applications of topological structures require the ability to manipulate them by using external stimuli and comprehensive understanding of their dynamic properties[13–17]. To date, extensive theoretical and experimental studies have been carried out to explore the evolution of topological structures under electric and mechanical fields[18–21]. For example, in $PbTiO_3$/$SrTiO_3$ superlattices, reversible phase transition between flux-closure arrays and trivial ferroelectric phase driven by either electric or mechanical fields have been observed[22]. In such a similar system, vortex arrays switch to out-of-plane and in-plane polarization by electric fields and mechanical loading, respectively[23,24]. However, in these studies, topological structures in the oxide superlattice always appear as arrays and thus the switching of them exhibits a collective behavior, usually involving multiple topological units. For instance, one clockwise vortex is always sandwiched between two anticlockwise ones in a PbTiO$_3$ layer[7], and similar for flux-closure domains[6]. Under external fields, they usually emerge and disappear simultaneously[22–24]. Therefore, the dynamic property of isolated topological polar structures under external stimuli is little known and the ability to control individual topological polar structures that is critical for practical applications, e.g., data storage, for which one-by-one writing and erasing is required, is still challenging.

[1]Electron Microscopy Laboratory, and International Center for Quantum Materials, School of Physics, Peking University, 100871 Beijing, China. [2]Department of Materials Science and Engineering, University of Toronto, Toronto, ON M5S 3E4, Canada. [3]Department of Materials Science and Engineering, The Pennsylvania State University, University Park, PA 16802, USA. [4]Beijing National Laboratory for Condensed Matter Physics, Institute of Physics, Chinese Academy of Sciences, 100190 Beijing, China. [5]Department of Materials Science and Engineering, National Yang Ming Chiao Tung University, Hsinchu 30010, Taiwan, ROC. [6]Department of Materials Science and Engineering, National Chung Hsing University, Taichung 40227, Taiwan, ROC. [7]School of Physical Sciences, University of Chinese Academy of Sciences, 100190 Beijing, China. [8]Songshan Lake Materials Laboratory, 523808 Dongguan, Guangdong, China. [9]Institute of Physics, Academia Sinica, Taipei 11529, Taiwan, ROC. [10]Collaborative Innovation Centre of Quantum Matter, 100871 Beijing, China. [11]Interdisciplinary Institute of Light-Element Quantum Materials and Research Center for Light-Element Advanced Materials, Peking University, 100871 Beijing, China. [12]Hefei National Laboratory, 230088 Hefei, China. ✉e-mail: p-gao@pku.edu.cn

In this work, we choose the threefold vertex as a model system to study dynamic properties of isolated topological polar structures. Threefold vertices are intersections of a 180° domain wall and two 90° domain walls in ferroelectrics[5,25] (Supplementary Fig. 1). They consist of smoothly rotated dipoles at the meeting point of domain walls, thus are considered as a type of topological polar structures[2,25]. The winding number which is employed to characterize topological structures[26,27], is calculated to be $+\frac{1}{2}$ for threefold vertices (Supplementary Fig. 1). Although threefold vertices have been observed before[4–6], the dynamic behavior of isolated ones under electric fields is largely unknown. The observation of isolated threefold vertices in ferroelectric films[4] suggests the possibility to explore dynamic properties of isolated ones while other topological polar structures such as vortices and polar skyrmions usually appear as arrays in (PbTiO$_3$)$_n$/(SrTiO$_3$)$_n$ superlattices[7,8]. However, the main challenge is that threefold vertices are usually generated on insulating substrates, for which the polarization charge at the interface is not completely screened[28]. Therefore, it is difficult to apply an electric field to study the dynamic behavior of threefold vertices due to the lack of the bottom electrode.

Here, we demonstrate the controlled formation and motion of isolated threefold vertices in a PbTiO$_3$ thin film on DyScO$_3$ with a SrRuO$_3$ buffer layer by an applied electric field. We elaborately use the atomic-thin diffusion layer SrTiO$_3$ to induce incomplete screening while remain the SrRuO$_3$ as the bottom electrode to achieve the application of an external electric field. Under electric fields, 180° domain walls reach the interface, leading to the formation of isolated threefold vertices to minimize the depolarizing field. We directly observe electric-field-driven motion of isolated threefold vertices along the interface in a controllable and reversible manner with a velocity up to ~629 nm s$^{-1}$. Phase-field simulations verify the role of the SrTiO$_3$ layer and reveal microstructural evolution details of the nucleation and propagation of isolated threefold vertices. These results elucidate the ability to electrically manipulate isolated threefold vertices and suggest the movable feature of isolated topological polar structures which provides useful information for applications.

## Results

### Formation and characterization of isolated threefold vertices

For this work, 100-nm-thick PbTiO$_3$ thin films were deposited on single-crystal (110)$_O$ DyScO$_3$ substrates (the subscript O indicates orthorhombic) with 50-nm-thick SrRuO$_3$ as the bottom electrode. Figure 1a shows a schematic of an isolated threefold vertex. A cross-sectional view of the PbTiO$_3$/SrRuO$_3$ interface is shown in Fig. 1b by a dark-field transmission electron microscopy (TEM) image. The typical a/c domain patterns and two vertical 180° domain walls are observed by diffraction contrast imaging. In the enlarged images in Fig. 1b, small triangular prisms with ~45° tilted domain walls indicate the existence of nanometer-scale threefold vertices at the interface. The atomic structure of isolated threefold vertices was obtained by an atomic-resolution high-angle annular dark-field (HAADF) image in Fig. 1c and further analyzed quantitatively[29,30]. One 180° domain wall and two 90° domain walls were denoted by white dashed lines. The out-of-plane and in-plane lattice parameters of threefold vertices were measured at the unit-cell scale and mapped in Fig. 1d and e, respectively (a careful calibration has been performed with DyScO$_3$ substrates as the reference). The spatial variation of lattice parameters across the threefold vertex are shown in Supplementary Fig. 2 corresponding to dashed boxes in Fig. 1d, e. The in-plane lattice parameters of triangular 90° domain is suppressed to ~405 pm compared with ~415.8 pm in typical 90° domains[31]. Besides the suppression of in-plane lattice parameters, the lattice rotation in the triangular 90° domain is also revealed by measuring the angle of the Pb sublattice in Fig. 1f. The contrast in the triangular prism indicates a wave-like lattice rotation. By wave-like lattice rotation and suppressed in-plane lattice parameters, the triangular prism smoothly fit into basic domains at atomic scale.

The formation of isolated threefold vertices is further verified by phase-field simulations through considering the properties of the interface as we discuss below. The simulated domain structure presented in Fig. 1g shows the formation of a pair of isolated threefold vertices at the intersections of the 180° domain walls and the PbTiO$_3$/SrTiO$_3$ interface. The maps of polarization arrangements of two isolated threefold vertices as well as their in-plane lattice deformation are shown in Fig. 1h, i, which agree well with our experimental measurements in Fig. 1e. The triangular 90° in-plane domain within each threefold vertex serves as a transition region between the downward and upward domains, which greatly reduces the dipole bound charges at the PbTiO$_3$/SrTiO$_3$ interface.

### Atomic structure of the PbTiO$_3$/SrTiO$_3$ interface

To figure out the underlying mechanism of why threefold vertices formed at a ferroelectric/electrode interface, we analyzed the atomic structure and elemental distribution of the interface. Figure 2a shows the atomic structure of the PbTiO$_3$/SrRuO$_3$ interface. The corresponding atomic-resolution energy-dispersive X-ray spectroscopy (EDS) mapping in Fig. 2b suggests the diffusion of Ti into SrRuO$_3$. The substitution of Ru by Ti generates an about two-unit-cell SrTiO$_3$ layer at the interface which is marked by dashed white lines. Since SrTiO$_3$ is insulating, it may impair the perfect screening of polarization charges at the interface. Figure 2c shows a map of relative displacement vectors of Ti cations relative to the center of surrounding Pb cations, which reflects the polarization distribution. A profile of polar displacements in Fig. 2d shows the suppression of polarization within about 6 unit cells from the interface while the transition layer is just ~1.5 unit cells at the PbZr$_{0.2}$Ti$_{0.8}$O$_3$/SrRuO$_3$ interface according to the previous study[32]. In addition, the displacement between cations even exists in SrTiO$_3$ and SrRuO$_3$. Therefore, the wider transition layer and displacements in SrTiO$_3$ and SrRuO$_3$ indicate the imperfect screening of depolarizing field induced by the diffused SrTiO$_3$ layer. The residual depolarizing field provides proper electrical conditions for the formation of threefold vertices[33–35]. This hypothesis is verified by phase-field simulations of a comparison between the polarization distributions in systems with and without a 0.8-nm-thick SrTiO$_3$ layer, as shown in Fig. 2e, f, respectively. As seen, the threefold vertex is only formed within the PbTiO$_3$ film possessing a SrTiO$_3$ transition layers on the top of the electrode, whereas the film without a SrTiO$_3$ layer demonstrates a sharp 180° domain wall without any additional polarization structures at the interface. The result proves the critical role of the SrTiO$_3$ layer, which is also consistent with previous studies of BiFeO$_3$ on insulating and conducting substrates[28].

### Electric-field-driven motion of isolated threefold vertices

Since the atomic-thin SrTiO$_3$ layer at the interface induces the formation of isolated threefold vertices, we elaborately employed this to study dynamic properties of isolated threefold vertices under electric fields by in situ TEM electrical testing system. The in situ TEM tests allow us to directly observe the domain structures at the interface and track their evolutions in ferroelectric films[36,37]. Figure 3a depicts the probe-based electrical testing system. A tungsten tip acts as the top electrode to apply a bias on PbTiO₃ films while the SrRuO₃ electrode is grounded. The distribution of electric fields in ferroelectric films has been simulated (Supplementary Fig. 3). A chronological diffraction contrast TEM image series in Fig. 3b–g shows the domain structure evolution under varying electric fields. Figure 3b presents the initial state of PbTiO₃ films with typical a/c domain patterns. The polarization in PbTiO₃ films is downward as illustrated in Fig. 2c. When a negative voltage is applied, a domain with upward polarization appears (Supplementary Movie 1). Nanoscale isolated threefold vertices formed where 180° domain walls meet the interface as shown in Fig. 3c. The threefold vertices are distinguished by small triangular domains with ~45° tilted domain walls as shown in enlarged images. When the voltage changes from −6 volts to zero, the isolated threefold vertices remain at the interface in Fig. 3d, indicating their stability.

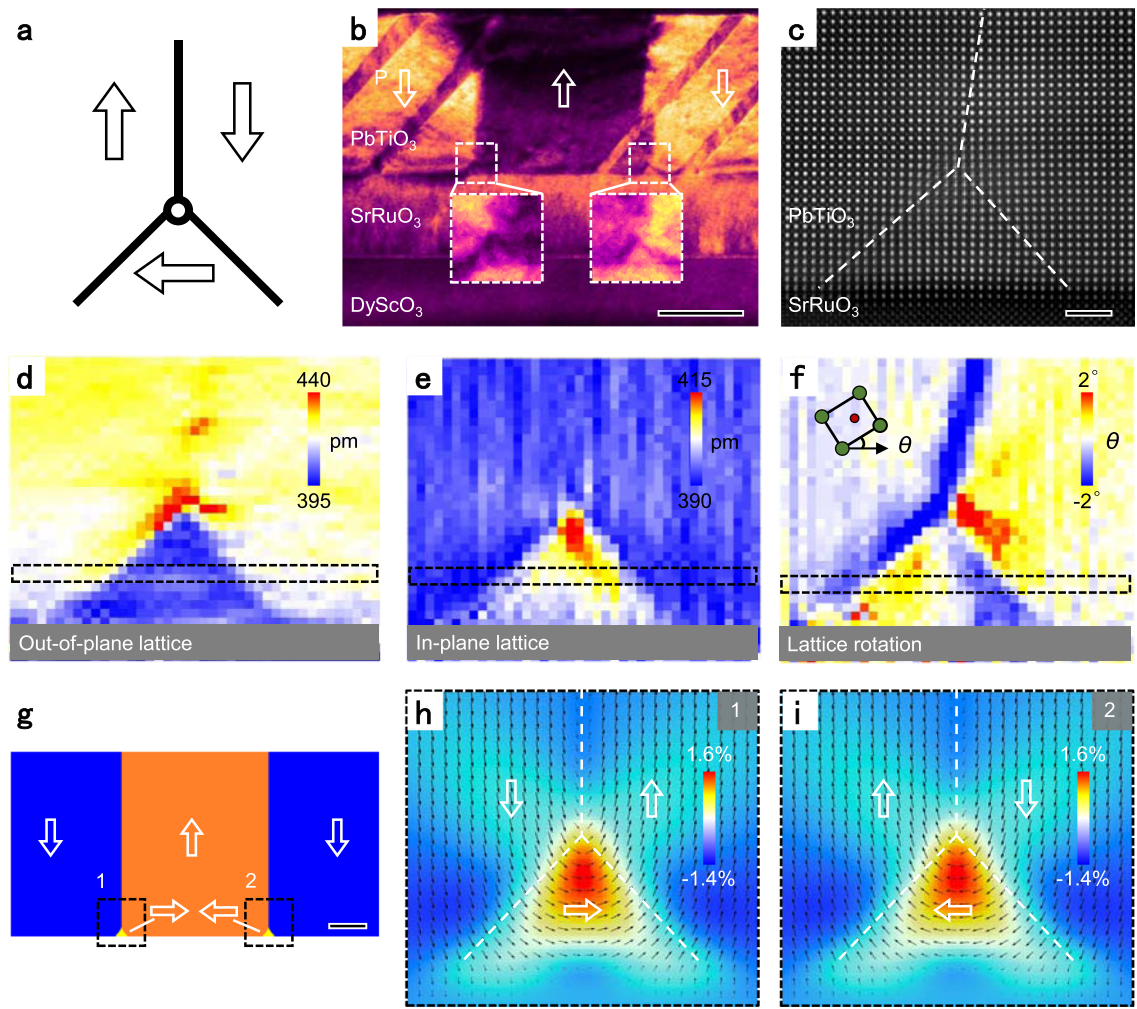

**Fig. 1 | Threefold vertices at the PbTiO₃/SrRuO₃ interface. a** Schematic diagram of a single threefold vertex. Arrows indicate directions of polarization. **b** A cross-sectional dark-field TEM image with **g** = (101) showing the domain structure in the PbTiO₃ film. The insets show the enlarged view of threefold vertices. Scale bar, 50 nm. **c** An atomically resolved HAADF image of an isolated threefold vertex at the PbTiO₃/SrTiO₃ interface. White dashed lines indicate 180° and 90° domain walls. Scale bar, 2 nm. **d–f** Maps of out-of-plane lattice parameters (**d**), in-plane lattice parameters (**e**), and lattice rotation (**f**). The inset schematic showing the definition of lattice rotation angle $\theta$. Profiles of lattice parameters across the threefold vertex are shown in Supplementary Fig. 2 corresponding to dashed boxes in **d–f**. **g** Formation of isolated threefold vertices from phase-field simulations. Scale bar, 20 nm. **h, i** Polarization configurations of the two isolated threefold vertices corresponding to dashed boxes in **g** with in-plane strain fields as backgrounds.

In addition, when the opposite electric field is applied, 180° domain walls move in the opposite direction in Fig. 3e and f. The nearly synchronous motion of 180° domain walls and threefold vertices along the interface can avoid a significant deviation of 180° domain walls from the symmetric permitted planes, because tilted 180° domain walls are charged and thus energetically unfavorable[38].

Finally, two threefold vertices coalesce and 180° domain walls leave the interface in Fig. 3g. We measured the distance $d$ between two isolated threefold vertices. In Fig. 3h, we plot $d$ and applied voltage $U$ as functions of time to quantitatively analyze the switching process. The distance between two isolated threefold vertices is up to ~134 nm, more than 10 times of their width (~10 nm). The velocity of isolated threefold vertices is in a range of 0–629 nm s⁻¹ as shown in Supplementary Fig. 4. Therefore, the generation and motion of threefold vertices is controlled by an applied voltage in a reversible manner. The creation and motion of isolated threefold vertices are repeatable (Supplementary Fig. 5 and Movie 2).

**Phase-field simulations of dynamic properties of isolated threefold vertices**

The nucleation and propagation of isolated threefold vertices in the PbTiO₃ film are reproduced by phase-field simulations in Fig. 4a–f. Starting from a downward polarization state same as the experiments, a voltage bias of a probe tip is applied at the top surface of the PbTiO₃ film. Under an increasingly negative voltage bias, a new domain with upward polarization nucleates near the tip, forming 180° domain walls (see Fig. 4b for the domain structure). The new domain grows vertically towards the substrate and then reaches the PbTiO$_3$/SrTiO$_3$ interface at −6.7 V. This is accompanied by an instant nucleation of a region of in-plane polarization between the bottom front of the 180° domain walls and the interface (Fig. 4b), with its polarization configuration presented in Fig. 4g. This leads to the formation of two isolated threefold vertices (Fig. 4c) whose polarization distribution are related to each other by a mirror symmetry as shown in Fig. 4h. The two isolated threefold vertices further move apart along the interface (Fig. 4d) under the negative voltage bias −8 V. The process can be reversed upon the application of an opposite voltage, in which the two isolated threefold vertices move towards each other (Fig. 4e) and finally coalesce, recovering to the original state of the PbTiO$_3$ film (Fig. 4f). The simulated nucleation and reversible motion of the isolated threefold vertices agree well with the experimental observations. Isolated threefold vertices do not vanish during the motion indicating their stability. The microstructure evolutions under electric fields not only verify the ability to move of isolated threefold vertices but also provide further details of the nucleation process.

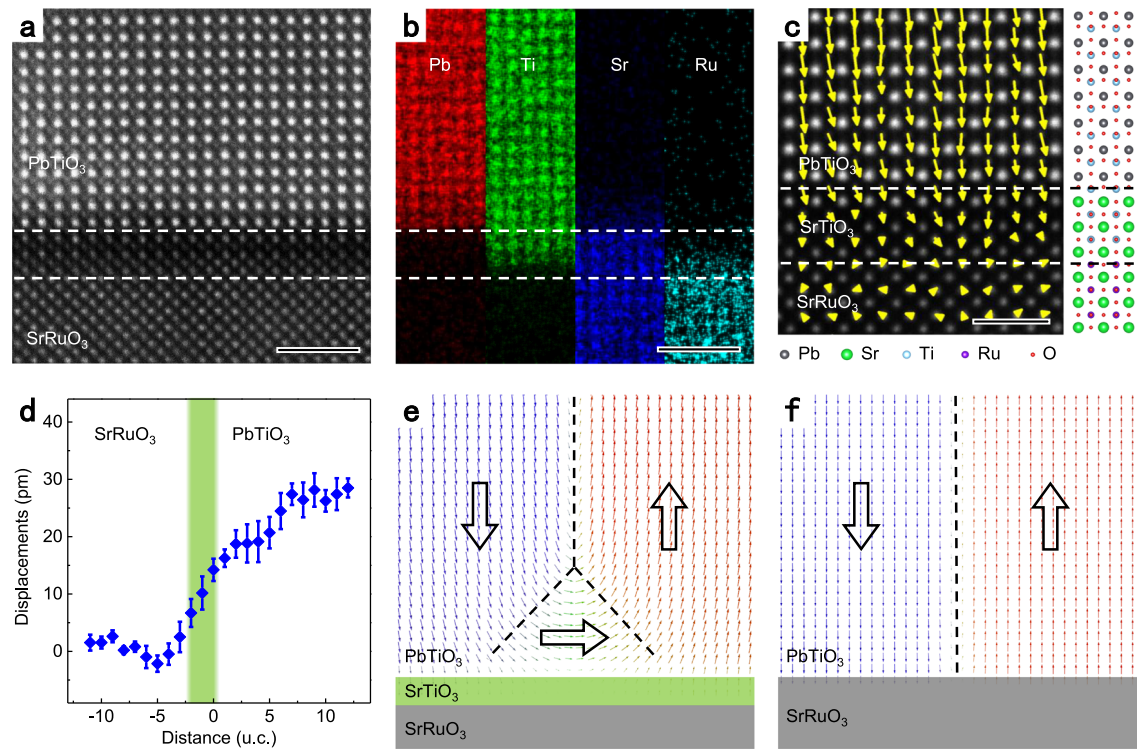

**Fig. 2 | Atomic structure and elemental diffusion at the PbTiO$_3$/SrRuO$_3$ interface. a** An atomically resolved HAADF image of the PbTiO$_3$/SrRuO$_3$ interface. Scale bar, 2 nm. **b** Corresponding atomic-resolution EDS elemental mappings for Pb, Ti, Sr, and Ru. White dashed lines indicate a two-unit-cell SrTiO$_3$ layer induced by Ti diffusion into SrRuO$_3$. Scale bar, 2 nm. **c** A mapping of polar displacements with a schematic diagram of the atomic structure of the interface. Scale bar, 1 nm. **d** The quantitative measurement of polar displacements as a function of the distance away from the PbTiO$_3$/SrRuO$_3$ interface. The colored region denotes the location of the diffused SrTiO$_3$ layer. Polarization configurations at interfaces with (**e**) and without (**f**) a SrTiO$_3$ layer from phase-field simulations. Arrows indicate directions of polarization.

## Discussion

For electric-field-induced threefold vertices at the interface, both electrical and mechanical boundary conditions are important. In electrical, the presence of a SrTiO$_3$ layer impairs perfect screening of metallic SrRuO$_3$ electrode. When 180° domain walls end at the interface, the formation of threefold vertices with triangular 90° domains can reduce polarization charges at the interface. Phase-field simulations further prove the critical role of SrTiO$_3$ layer for electric-field-induced threefold vertices. From a mechanical perspective, a tensile strain from DyScO$_3$ substrates provides proper mechanical boundary conditions for the formation of threefold vertices. The in-plane lattice mismatch between DyScO$_3$ and PbTiO$_3$ is 1.3%, calculated by the following formula[39]:

$$f = \frac{2(a_2 - a_1)}{(a_2 + a_1)} \times 100\% = 1.3\% \quad (1)$$

where $a_1$ is the in-plane lattice parameter of PbTiO$_3$ films, 390.4 pm, and $a_2$ is the in-plane lattice parameter of DyScO$_3$, 395.6 pm[40]. We also characterized the atomic structure of 180° domain walls at the PbZr$_{0.2}$Ti$_{0.8}$O$_3$/SrTiO$_3$ interface for comparison (Supplementary Fig. 6). The in-plane lattice mismatch between PbZr$_{0.2}$Ti$_{0.8}$O$_3$ and SrTiO$_3$ is negligible, i.e., 390.4 pm in PbZr$_{0.2}$Ti$_{0.8}$O$_3$ and 390.5 pm in SrTiO$_3$[41]. Without the tensile stress, a sharp 180° domain wall appears at the PbZr$_{0.2}$Ti$_{0.8}$O$_3$/SrTiO$_3$ interface without threefold vertices (Supplementary Fig. 6). Therefore, the two-unit-cell-thick SrTiO$_3$ layer not only mediate the screening of depolarizing field but also maintain the strain field from the substrate. These results demonstrate the possibility of employing interface engineering to fabricate isolated topological polar structures by applied electric fields. Their isolated nature can be verified by the geometric phase analysis (GPA) and motion analysis. From GPA results (Supplementary Fig. 7), the strain field of these two threefold vertices is very local ~10 nm, which is much less than their distance (over 100 nm), indicating no stress interaction between them. We also found that these two isolated threefold vertices can move independently (Supplementary Fig. 8), which provides additional evidence for their isolation.

The dynamic property of isolated structures is critical as they may be regarded as single functional elements. Many impressive progress has been made in exploring dynamic behavior of topological structures by state-of-art electron microscopy and other techniques[20–22,42,43]. During the collective switching process, these

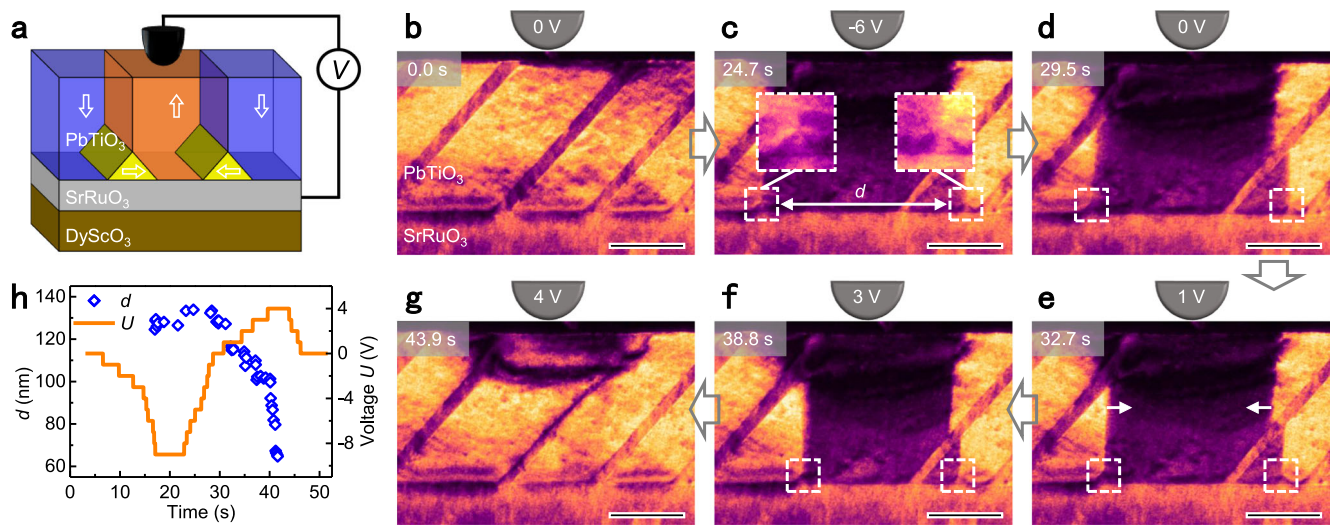

**Fig. 3 | Electric-field control of isolated threefold vertices at the PbTiO₃/SrRuO₃ interface. a** Schematic diagram of the probe based in situ TEM experimental set-up. Two isolated threefold vertices are formed at the PbTiO₃/SrRuO₃ interface. Arrows indicate directions of polarization. **b–g** A chronological TEM dark-field image series illustrates the nucleation and lateral motion of two threefold vertices along the PbTiO₃/SrRuO₃ interface under applied electric fields. White arrows in **e** indicate directions of the domain wall motion. Scale bar, 50 nm. **h** Plots of applied voltage (orange line) and distance between two isolated threefold vertices (blue diamond) as functions of time.

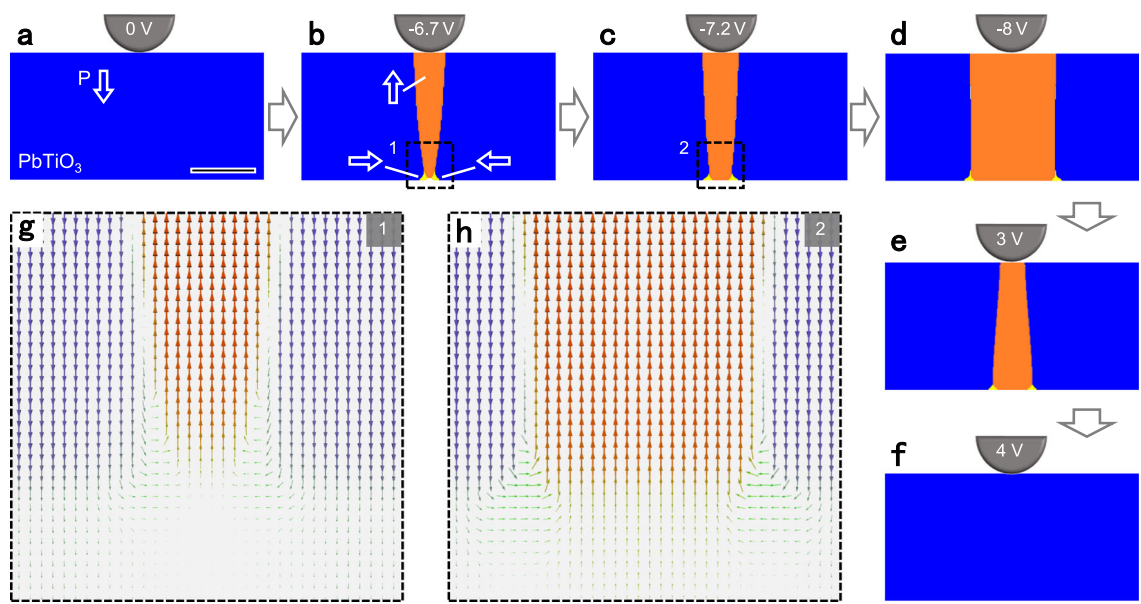

**Fig. 4 | Phase-field simulations of electric-field control of nucleation and motion of isolated threefold vertices. a** The initial state of the PbTiO₃ film with downward polarization. A voltage bias can be applied at the top surface of the PbTiO₃ film through a metal tip. White arrow indicates the direction of polarization. Scale bar, 50 nm. **b** The isolated threefold vertices nucleate when the upward domain reaches the interface. **c, d** Isolated vertices move away from each other under a negative voltage. **e** Isolated vertices move towards each other under a positive voltage. **f** The PbTiO₃ film returns to its initial polarization state. **g** The polarization configuration of the nucleation region corresponding to the dashed box in **b**. **h** The polarization configuration of electric-field-induced threefold vertices corresponding to the dashed box in **c**.

topological structure arrays are almost fixed at their locations in superlattices. In contrast, our results demonstrate the ability to move of isolated threefold vertices which enables the design of devices such as electric racetrack memory[12,44]. The physical origin of the ability to move of isolated threefold vertices can be understood from the energy perspective. We carried out phase-field simulations to compare the energy of systems with different domain wall configurations in Supplementary Fig. 9. The system with a threefold vertex has a lower energy compared with the system without a threefold vertex, which is reasonable as the formation of threefold vertices reduces the amount of polarization charges at the interface[4,28]. In other words, the 180° domain walls and threefold vertices are strongly correlated in this system. Under electric fields, the 180° domain wall moves to increase the domain volume with a polarization parallel to the external electric field[15,45]. If the corresponding threefold vertex do not move, the 180° domain wall will tilt and deviate from the symmetric permitted plane. For such a tilted 180° domain wall configuration, the calculated system energy becomes much higher than that with a vertical 180° domain wall (i.e., perpendicular to the interface) as shown in Supplementary

Fig. 9. Therefore, it is energetically favorable for the isolated threefold vertex to move together with the moving 180° domain wall.

The ability to move of isolated threefold vertices suggests the possible movable feature of isolated topological polar structures. However, such dynamic behavior is still largely unexplored compared with their counterparts in magnetics. In magnetics, many isolated topological structures have been proven to be movable and controllable under external stimuli. For example, isolated magnetic skyrmions can be generated and driven by current[46]. The skyrmion Hall effect has also been discovered by driving the motion of isolated magnetic skyrmions[47]. In situ TEM combined with interface engineering offers an effective strategy to probe the dynamic properties of isolated topological polar structures.

In summary, our work demonstrates electric-field-controlled formation and motion of isolated threefold vertices. We elaborately use the atomic-thin diffusion layer SrTiO$_3$ to induce incomplete screening that is prerequisite to generate threefold vertices at the PbTiO$_3$/SrRuO$_3$ interface, and meanwhile achieve the application of an external electric field. These isolated threefold vertices are movable under electric fields in a controllable and reversible manner with a velocity up to ~629 nm s$^{-1}$. The nucleation and long-range motion of isolated threefold vertices are reproduced by phase-field simulations. This work demonstrates the possibility of manipulating isolated topological structures via interface engineering and electric fields. The ability to move of nanoscale threefold vertices provides insights into the dynamic property of isolated topological polar structures and possibilities for applications.

## Methods
### Film growth
The 100-nm-thick PbTiO$_3$ thin films were grown on single crystal (110)$_o$ DyScO$_3$ substrates buffered with ~50-nm-thick SrRuO$_3$ electrodes using pulsed laser deposition method, as well as the PbZr$_{0.2}$Ti$_{0.8}$O$_3$ films on (001) SrTiO$_3$ substrates. A KrF excimer laser ($\lambda$ = 248 nm) was focused on the targets with an energy density of ~2.5 mJ cm$^{-2}$ and repetition rate of 10 Hz. During the growth of the SrRuO$_3$, the temperature and the oxygen pressure were kept at 700 °C and 100 mTorr, respectively. To avoid the easy evaporation of the lead in PbTiO$_3$ thin film at high temperature, the target with 5% excess lead was used and the temperature for the subsequent growth of PbTiO$_3$ thin film was decreased to 650 °C at the same oxygen atmosphere. After the deposition process, an in situ post annealing process with the same temperature and high oxygen pressure environment of ~300 Torr for 30 min was adopted to effectively eliminate the oxygen vacancies of the sample. Finally, the sample was slowly cooled down to room temperature with a ramp rate of 5 °C min$^{-1}$.

### Sample preparation and characterization
TEM samples were prepared by conventional mechanical polishing and subsequent argon ion milling in a Precision Ion Polishing System 691 (Gatan). The procedure for ion milling consisted of two steps. During the first stage, the guns were operated at 4 keV and at angles of 6° and −6°. During the second stage, the guns were operated at 1 keV for 5 min and at angles of 3° and −3°, and lowered further to 0.1 keV for 2 min for final surface cleaning. Diffraction contrast TEM experiments were carried out using a FEI Tecnai F20 microscope. Samples were tilted off the zone axis and imaged in the bright field or dark field with two-beam alignment condition using a **g** = (101) vector. High-angle annular dark-field (HAADF) images and energy-dispersive X-ray spectroscopy (EDS) images in this work were obtained using probe aberration-corrected FEI Titan Cubed Themis G2 operated at 300 kV in Electron Microscopy Laboratory of Peking University. Atom positions were determined by simultaneously fitting two-dimensional Gaussian peaks to a perovskite unit cell using a home-developed code. Polar displacements of the Ti cations were measured relative to the center of the surrounding Pb cations in HAADF images. Annular bright field (ABF) images were recorded at 300 kV in JEM ARM300CF (JEOL Ltd.). The convergence semi-angle for imaging is 24 mrad, collection semi-angles snap is 12–24 mrad for ABF.

### In situ study
In situ TEM experiments were carried out on a FEI Tecnai F20 microscope operated at 200 kV with a PicoFemto double-tilt TEM-STM holder provided by ZEPTools Technology Company. A tungsten tip acted as the top electrode, which was precisely controlled by a piezoelectric system. The switching processes were recorded with a OneView camera (Gatan). The imaging rate was set at 10 frames per second in order to get a good contrast.

### Phase-field simulations
In the phase-field model, the time evolution of the polarization field **P**(**x**) and mechanical displacement field **u**(**x**) are described by the time-dependent Ginzburg–Landau equation and the elastic equilibrium equation, respectively[48], i.e.,

$$\frac{\partial \mathbf{P}}{\partial t} = -L_P \frac{\delta F}{\delta \mathbf{P}} \tag{2}$$

$$\frac{\delta F}{\delta \mathbf{u}} = 0 \tag{3}$$

F[**P**, **u**] is the total free energy of the system formulated as a function of the polarization field **P** and the mechanical displacement field **u** with expressions the same as those in ref. 49. $L_P$ is the kinetic coefficient of polarization. A semi-implicit Fourier-spectral method[50] is employed for the numerical simulations.

The simulation system of the ferroelectric film consists of 4 layers, including a 4-nm-thick vacuum top layer, a 100-nm-thick PbTiO$_3$ film layer, a 0.8-nm-thick SrTiO$_3$ transition layer, and a 35.2-nm-thick substrate layer (i.e., the total vertical size of the system is 140 nm). The in-plane size of the system is taken as 200 nm. An electrostatic boundary condition of a specified voltage applied by the probe tip is employed at the top surface of the PbTiO$_3$ film[51]. A grounded electrostatic boundary condition is employed at the top plane of the substrate layer (which connects to the bottom of the SrTiO$_3$ layer). For the separate study of a film without a SrTiO$_3$ transition layer, the PbTiO$_3$ layer and the substrate layer are directly connected, where a grounded electrostatic boundary condition is employed at their interface. The material constants of PbTiO$_3$ and SrTiO$_3$ adopt the same values as those in ref. 52.

### Winding number calculations
The winding number is contour integral of the variation of polarization orientation along a closed loop. The winding number $n$ of threefold vertices is calculated with the following equation[27]:

$$n = \frac{1}{2\pi} \oint_C \Delta\theta \cdot dr \tag{4}$$

where $\theta$ is polarization orientation measured anticlockwise from horizontal direction. $\Delta\theta$ is the orientation difference between of two neighboring points. The winding number of the threefold vertex is $+\frac{1}{2}$. We assume that 180° domain wall has no contribution because of collinear polarization vectors[53]. A flux-closure structure (a pair of threefold vertices) has a winding number +1, which is equal to the sum of these two threefold vertices. This agrees with the conservation property of winding numbers[27].

**Reporting summary**

Further information on research design is available in the Nature Research Reporting Summary linked to this article.

## Data availability

The data that support the findings of this study are available within the article and the Supplementary Information. Any other relevant data are also available upon reasonable request from the corresponding author.

## Code availability

Additional data including the codes are available from the corresponding author upon reasonable request.

## Acknowledgements
P.G. is supported by the National Natural Science Foundation of China (52125307, 52021006). The phase-field simulations by T.Y. and L.-Q.C. are supported as part of the Computational Materials Sciences Program funded by the U.S. Department of Energy, Office of Science, Basic Energy Sciences, under Award No. DE-SC0020145. The phase-field simulations were performed on the Extreme Science and Engineering Discovery Environment (XSEDE) supported by National Science Foundation grant number ACI-1548562.

## Author contributions
M.L. carried out TEM experiments, analyzed the data and wrote the manuscript under the direction of P.G.; T.Y. and L.-Q.C. carried out phase-field simulations. P.C., X.B., X.M., and J.Z. assisted with TEM characterization. Y-J.W., H.-J.L., and Y.-L.H. grew the samples under the direction of Y.C.; R.Z., X.L., and R.S. assisted with data analysis. All authors contributed to this work through useful discussion and comments to the manuscript.

## Competing interests
The authors declare no competing interests.

## Additional information
**Supplementary information** The online version contains supplementary material available at https://doi.org/10.1038/s41467-022-33973-8.

**Correspondence** and requests for materials should be addressed to Peng Gao.

**Peer review information** *Nature Communications* thanks Vincent Garcia and the other anonymous reviewer(s) for their contribution to the peer review of this work.

**Reprints and permission information** is available at http://www.nature.com/reprints

**Publisher's note** Springer Nature remains neutral with regard to jurisdictional claims in published maps and institutional affiliations.